\documentclass[aps,pre,preprint,superscriptaddress,floatfix]{revtex4-1}
\usepackage{bm}
\usepackage{graphicx,color}
\usepackage{graphicx}
\usepackage{dcolumn}
\usepackage{times}

\usepackage{amsmath,amssymb}

\begin{document}

\title{Whodunnit? The case of midge swarms} 
\author{L. L. Bonilla$^*$}
\affiliation{Gregorio Mill\'an Institute for Fluid Dynamics, Nanoscience and Industrial Mathematics, Universidad Carlos III de Madrid, 28911 Legan\'{e}s, Spain}
\affiliation{Department of Mathematics, Universidad Carlos III de Madrid, 28911 Legan\'{e}s, Spain.\\
$^*$Corresponding author. E-mail: bonilla@ing.uc3m.es. ORCID: 000-0002-7687-8595}
\author{R. Gonz\'alez-Albaladejo$^{**}$}
\affiliation{Sorbonne Universit\'e, Laboratoire de Physique Th\'eorique et Hautes Energies, CNRS, UMR 7589, 4 Place Jussieu, 75252 Paris Cedex 05, France}
\affiliation{Gregorio Mill\'an Institute for Fluid Dynamics, Nanoscience and Industrial Mathematics, Universidad Carlos III de Madrid, 28911 Legan\'{e}s, Spain.\\
$^{**}$ORCID: 0000-0001-9560-5720}
\date{\today}
\begin{abstract}
As collective states of animal groups go, swarms of midge insects pose a number of puzzling questions. Their ordering polarization parameter is quite small and the insects are weakly coupled among themselves but strongly coupled to the swarm. In laboratory studies (free of external perturbations), the correlation length is small, whereas midge swarms exhibit strong correlations, scale free behavior and power laws for correlation length, susceptibility and correlation time in field studies. Data for the dynamic correlation function versus time collapse to a single curve only for small values of time scaled with the correlation time. Is there a theory that explains these disparate observations? Among the existing theories, whodunnit? Here we review and discuss several models proposed in the literature and extend our own one, the harmonically confined Vicsek model, to anisotropic confinement. Numerical simulations of the latter produce elongated swarm shapes and values of the static critical exponents between those of the two dimensional and isotropic three dimensional models. The new values agree better with those measured in natural swarms.
\end{abstract}
\maketitle


\section{Introduction}\label{sec:1}
For animals of such modest size, swarms of midge insects pose a number of puzzling questions. Unlike bird \cite{bal08} and sheep \cite{gin15} flocks, fish schools \cite{her11} or human crowds \cite{kar14}, the polarization alignment of midge swarms is always quite small and the associated patterns rather simple \cite{att14plos}. Despite this, natural swarms exhibit strong correlations and scale free behavior: the correlation length is proportional to the size of the swarm \cite{att14plos,att14}. However, swarms formed in the laboratory under controlled conditions in the absence of external perturbations are weakly coupled and have a small correlation length. Adding perturbations restores the scale free property \cite{vaa20}. Having scale free behavior, magnitudes related to correlation functions exhibit power laws \cite{att14,cav17}.

Power laws seem to be ubiquitous in nature \cite{mor11,bia12} but how can be sure they are for real? Stumpf {\em et al} have proposed that a power law has to exhibit a linear behavior in a log-log plot for at least two orders of magnitude, there should be a theory that generates it, and there should be ample and uncontroversial empirical support for it \cite{stu12}. Equilibrium second order phase transitions (e.g., paramagnetic to ferromagnetic phases at the Curie temperature \cite{hua87,ami05}) satisfy all these requirements and Wilson's renormalization group ideas about universality have been enormously fruitful to explain many phenomena beyond the original discoveries of power laws \cite{wil74,wil83,hoh77}. How useful are these ideas to explain midge swarming?

Here we explore midge swarming seeking to explain the known empirical facts that we take as clues. Similar to popular murder cases ``Whodunnit?'', we examine different theories (the suspects) that fit some of the facts and discard those that fail to explain them all, thereby pointing out the most likely suspect. Section \ref{sec:2} presents known features of swarms from experiments in the lab and from field studies. Four possible theories are briefly presented and commented upon in Section \ref{sec:3}. Section \ref{sec:4} discusses the theory that fits most clues and extends it to cover the anisotropic case hence providing new results. Then Section \ref{sec:5} discusses our findings, gives an overall picture of the subject and suggests possible future research.

 \section{Experiments and observations}\label{sec:2}
 Observation of insect swarming occur in nature and in the laboratory under controlled conditions. In nature, male midges form swarms at dawn or dusk over distinctive spots on the ground (wet areas, cow dung, man-made objects, etc) called markers \cite{dow55,oku74}. Their purpose is to attract females. Since the 2010s, Ouellette and coworkers have studied swarm formation and properties under controlled laboratory conditions \cite{oue22}. Environmental illumination levels provide a behavioral cue to individual insects for when to swarm. Controlled illumination show that swarm formation and dissolution is clearly reflected in the rapid establishment and disappearance of an emergent central potential that binds individuals to the swarm \cite{pat22}. Swarms consist of a condensed core and a vapor of insects that leave or enter it \cite{sin17}. Midges acoustically interact when their distances are sufficiently small \cite{att14plos,att14} and react collectively to external acoustic signals \cite{ni15}. The distribution of speeds is peaked about some value and exhibits heavy tails for large swarms (perhaps due to the formation of clusters) \cite{kel13}. The statistics of accelerations of individual midges in a swarm is consistent with postulating a linear spring force (therefore a harmonic potential) that binds insects together \cite{kel13}. The trajectories of single midges follow L\'evy flights \cite{rey16}. The polarization order parameter of the swarm,
 $W\in[0,1]$: 
\begin{eqnarray}
W(t)= \left|\frac{1}{N}\sum_{j=1}^N \frac{\mathbf{v}_j(t)}{|\mathbf{v}_j(t)|}\right|\!,\quad W=\langle W(t)\rangle_t,\label{eq1}
\end{eqnarray}
 is very small \cite{att14plos}. Here $\mathbf{v}_j(t)$ is the  velocity of the midge $j$, $j=1,\ldots,N$, at time $t$ and $W$ is the time average over the duration of the film taken from the observation of the swarm.
 
 In nature, midge velocities within swarms are strongly correlated and scale free, i.e., the correlation length is proportional to the size of the swarm \cite{att14plos,att14,cav17}. In contrast to field studies, swarms exhibit only short range velocity correlations in the laboratory, which are due to controlled temperature and absence of external environmental effects (breeze, temperature gradients, fluid flow, etc) \cite{niou15}. When perturbations are added, the correlation length is proportional to swarm size \cite{vaa20}. Field studies by Cavagna and coworkers have shown that the static and dynamic velocity correlation functions have peculiar qualitative features \cite{cav17}.  In critical dynamics about equilibrium second order phase transitions \cite{hoh77}, curves of the dynamic correlation function versus time collapse to a single curve as they are written as functions of time scaled with the correlation time (or, equivalently, with correlation length to a power given by the dynamic critical exponent). Field data show that the collapse of the dynamic correlation function with rescaled time occurs only for short times \cite{cav17}. In addition, the dynamic correlation function versus scaled time is flat at the origin in the sense defined in Ref.~\cite{cav17}. Furthermore, quantities such as the susceptibility, the correlation length and the correlation time follow power laws with characteristic static \cite{att14} and dynamic \cite{cav17,cav23} exponents. 
 
To be more precise, we need a few definitions. The static correlation function (SCF) $C(r)$ is  \cite{att14}
\begin{subequations}\label{eq2}
\begin{eqnarray}
&&C(r)= \frac{d Q(r)}{d r},\quad  Q(r)= \left\langle\frac{1}{N}\sum_{i=1}^{N}\sum_{j\neq i}^{N} \delta\hat{\mathbf{v}}_i(t)\!\cdot\!\delta\hat{\mathbf{v}}_j(t)\theta(r-r_{ij}(t))\right\rangle_t, \label{eq2a}\\
&&\delta\mathbf{v}_j(t)=\mathbf{v}_j(t)-\frac{1}{N}\sum_{l=1}^N \mathbf{v}_l(t),\quad \delta\hat{\mathbf{v}}_j(t)=\frac{\delta\mathbf{v}_j(t)}{\sqrt{\frac{1}{N}\sum_{k=1}^N|\delta\mathbf{v}_k(t)|^2}},\quad r_{ij}(t)=|\mathbf{x}_i(t)-\mathbf{x}_j(t)|,\quad\label{eq2b}
\end{eqnarray}\end{subequations}
where $\theta(x)=1$ for $x>0$ and $\theta(x)=0$ otherwise is the unit step function. The insect positions $\mathbf{x}_j(t)$ are measured from the center of mass and $Q(r)$ is the static cumulative correlation function. The Fourier transform of the dynamic correlation function (DCF) is \cite{cav17}
\begin{eqnarray}
\hat{C}(k,t)\!=\!\left\langle\! \frac{1}{N}\sum_{i,j =1}^{N}\!\frac{\sin(kr_{ij}(t_0,t))}{kr_{ij}(t_0,t)}\delta\hat{\mathbf{v}}_i(t_0)\!\cdot\!\delta\hat{\mathbf{v}}_j(t_0+t) \!\right\rangle_{t_0}\!,\quad \langle f\rangle_{t_0}=\frac{1}{t_{max}-t}\sum_{t_0=1}^{t_{max}-t}f(t_0,t),      \label{eq3}
\end{eqnarray}
where $r_{ij}(t_0,t)=|\mathbf{x}_i(t_0)-\mathbf{x}_j(t_0+t)|$, $k$ is the wave number and $t_0$ and $t_{max}$ are the initial and maximum observation time, respectively. From the SCF, the correlation length $\xi=r_0$ and the susceptibility $\chi$ are 
\begin{eqnarray}
C(r_0)=0,\quad \chi=\max_r Q(r)=Q(r_0).\label{eq4}
\end{eqnarray}
 In a second order equilibrium phase transition, the free energy is a homogeneous function of control parameter (e.g., temperature minus critical temperature), external field and correlation length. Differentiation of the free energy provides power laws for correlation length, susceptibility, etc in terms of the control parameter thanks to the homogeneity property, which is called finite-size scaling (FSS) \cite{ami05}. In field studies of midge swarms, FSS is assumed and the role of the temperature in the control parameter is played by the average distance $x$ between next neighbor insects (rescaled by the body length of a midge), which is called perception range \cite{att14}. Then the following power laws have been established
 \begin{eqnarray}
x-x_c\sim N^{-\frac{1}{3\nu}},\quad\xi\sim N^{1/3}\sim(x-x_c)^{-\nu},\quad\chi\sim(x-x_c)^{-\gamma}\sim N^\frac{\gamma}{3\nu}, \label{eq5}
\end{eqnarray}
where the measured static critical exponents are $\nu=0.35\pm 0.1$, $\gamma= 0.9\pm 0.2$. The value of $x$ for finite $N$ is the argument at which the susceptibility reaches its maximum and $x_c=12.5\pm 1.0$ ($N=\infty$). For critical dynamics about a second order equilibrium phase transition, the curves $\hat{C}(k,t)$ collapse into a single curve when the time is scaled with the correlation length to the dynamic critical exponent $z$, which is called the {\em dynamic scaling hypothesis} \cite{hoh77}. Assuming this hypothesis to hold, we could write Eq.~\eqref{eq3} as:
\begin{subequations}\label{eq6}
\begin{eqnarray}
&&\frac{\hat{C}(k,t)}{\hat{C}(k,0)}= \left. f\!\left(\frac{t}{\tau_k},k\xi\right)\right|_{k\xi=1}=g(k^zt,k\xi) \quad (k\propto \xi^{-1}),\label{eq6a}\\
&&g(t)=\frac{\hat{C}(k_c,t)}{\hat{C}(k_c,0)};\quad \tau_k=k^{-z}\phi(k\xi). \label{eq6b}
\end{eqnarray}
Here $\tau_k=k^{-z}\phi(k\xi)$ is the correlation time at wave number $k$, found by solving the equation \cite{cav17,gon23,gon24}:
\begin{eqnarray}
\sum_{t=0}^{t_{max}} \frac{1}{t}\,\sin\!\left(\frac{t}{\tau_k}\right) f\!\left(\frac{t}{\tau_k},k\xi\right)\! = \frac{\pi}{4}.  \label{eq6c}
\end{eqnarray}
\end{subequations}
The value at which $\hat{C}(k,t)$ reaches its maximum, $k_c$, is inversely proportional to the correlation length (so that we can redefine $\xi=1/k_c$, hence $\tau_{k_c}= \phi(1) \xi^z$). Note that the perception range is not necessary to calculate the dynamic critical exponent $z$. However, its value depends on the  method used to fit the data. Least square (LS) fitting yields $z=1.16\pm 0.12$ whereas reduced major axis (RMA) fitting produces $z=1.37\pm 0.11$ \cite{cav23}. With data obtained from field studies, the collapse into a single function $g(x)$ as given by Eq.~\eqref{eq6a} does not occur: All $\hat{C}(k,t)$ curves collapse into a single one but only on the interval $0<k_c^zt<4$ and they clearly do not overlap for larger values of $k_c^zt$ (cf Fig.~2 of Ref.~\onlinecite{cav17}). Besides partial collapse of the DCF, the function $-x^{-1}\ln g(x)$ tends to 0 (it is flat) as $x=k_c^zt\to 0$ \cite{cav17}, which is the other qualitative feature mentioned before.
 
 It is interesting to note that the data used to get the power laws and critical exponents for natural swarms were collected in different days over many years from swarms having different number of insects, and there were swarms of different midge species ($N$ ranges from 69 to 781; see Supplementary Materials of Refs.~\onlinecite{att14,cav17,cav23}).
 
 \section{Theories}\label{sec:3}
Here we present different theories of swarm formation and discuss whether they explain the observations described in Section \ref{sec:2}. We mostly focus on observations of natural swarms.

 \subsection{Periodic Vicsek model}
The periodic Vicsek model (PVM) \cite{vic95,vic12} consists of the equations
\begin{eqnarray}
&&\mathbf{x}_j(t+\Delta t)=\mathbf{x}_j(t)+ \Delta t\,\mathbf{v}_j(t+\Delta t),\quad j=1,\ldots,N,\nonumber\\
&& \mathbf{v}_j(t+\Delta t)=v  \mathcal{R}_\eta\!\left[\Theta\!\left(\sum_{|\mathbf{x}_l-\mathbf{x}_j|<r_1 R_0}\mathbf{v}_l(t)\right)\!\right]\!. \label{eq7}
\end{eqnarray}
Here $\Theta(\mathbf{x})=\mathbf{x}/|\mathbf{x}|$, $R_0$ is the radius of the sphere of influence about particles, $v_0$ is the constant particle speed, and $\mathcal{R}_\eta(\mathbf{w})$ performs a random rotation uniformly distributed on a spherical sector around $\mathbf{w}$ with maximum opening $\eta$ \cite{att14}. Particles align their velocities with the mean of their neighbors within a sphere of influence except for an alignment noise of strength $\eta$ and are confined to a cube of finite volume by periodic boundary conditions. As $v\Delta t/r_1$ tend to zero, the discrete time equations \eqref{eq7} become differential equations that describe point particles undergoing overdamped dynamics with forces that tend to align their velocities.

In 2D, the PVM exhibits an ordering phase transition from a spatially homogeneous phase (zero polarization, $W=0$) to traveling bands ($W>0$) \cite{cha20}. For a sufficiently small box length (much smaller than the period of the traveling bands), this ordering transition is continuous. Thus it would seem that this model is a plausible candidate to explain swarm formation if studied on the disordered phase near the phase transition (leaving aside the fact that observed swarms are far from being spatially homogeneous). However, Cavagna {\em et al} have shown that this is not the case: The DCF $g(k_c^zt)$ is not flat at the origin ($-x^{-1}\ln g(x)\to 1$ as $x\to 0$) and the numerically calculated critical exponents from the model are far from the measured ones (e.g., $z=2$, far from 1.37) \cite{cav17}. 

\subsection{Inertial spin models}
The inertial spin model (ISM) consists of the equations \cite{cav15,cav24}
\begin{eqnarray}
&&\frac{d\mathbf{x}_j}{dt}=\mathbf{v}_j,\quad j=1,\ldots,N,\nonumber\\
&&\frac{d\mathbf{v}_j}{dt}=\frac{1}{\chi}\mathbf{s}_j\times \mathbf{v}_j,\nonumber\\
&&\frac{d\mathbf{s}_j}{dt} =\mathbf{v}_j\times \frac{J}{n_i} \sum_{l=1}^N n_{jl}(t) \mathbf{v}_l -\frac{\eta}{\chi}\mathbf{s}_j - \mathbf{v}_j\times \boldsymbol{\zeta}_j, \label{eq8}
\end{eqnarray}
to be solved with periodic boundary conditions. Here the zero-mean white noise has correlation $\langle \zeta^\mu_j(t) \zeta^\nu_l(t')\rangle = 2T \eta\delta_{jl}\delta_{\mu\nu}\delta(t - t')$, $n_{jl}=\theta(R_0-|\mathbf{x}_j-\mathbf{x}_l|)$, $n_j=\sum_{l=1}^Nn_{jl}$. The parameter $J$ is the alignment strength, $\eta$ is the microscopic spin dissipation, $\chi$ is a spin scale, and $T$ is the noise amplitude (or temperature). Clearly the speed of the particles is conserved by these equations so that we can impose that they all have the same speed $v_0$. In the overdamped limit, we may ignore $d\mathbf{s}_j/dt$ and obtain the following equation:
\begin{eqnarray*}
\eta\frac{d\mathbf{v}_j}{dt}=\!\left(\mathbf{v}_j\times \tilde{J} \sum_{l=1}^N n_{jl}(t) \mathbf{v}_l +\boldsymbol{\zeta}_j\right)\!\times\mathbf{v}_j,
\end{eqnarray*}
with $\tilde{J}$ instead of $J/n_j$. This is a version of the continuous time Vicsek model and it has an ordering transition. Eq.~\eqref{eq8} adds inertia so that the resulting ISM may produce a DCF which is flat at the origin near the ordering transition \cite{cav15,cav24}. 

It turns out that the periodic ISM near its ordering transition produces the same dynamic critical exponent as the active irreversible version of model G in Ref.~\cite{hoh77}. The resulting model nonlinearly couples a field of unit velocity vectors with a spin field and it includes space-time white noise sources. At the ordering transition, Cavagna {\em et al} have found the renormalization group equations (RGEs) of the active G model within the one-loop approximation by using the $4-d$ Wilson expansion \cite{cav23}. The authors consider that one of the four critical points of the RGEs corresponds to the behavior of midge swarms and proceed to find the critical exponents using the most appropriate fixed point with the result: $z=1.35$, $\nu = 0.748$, $\gamma = 1.171$. Numerical solution of the ISM produces the same exponent $z$ and so it does a version of the ISM (fully-conserved ISM) that is closer to the active model G \cite{cav24}. Thus, active G model and the different ISMs may belong to the same universality class in Wilson's sense \cite{wil74}.

While the dynamical critical exponent predicted by these theories is close to observations using the RMA fit \cite{cav23}, the predicted static exponents are far from those measured ($\nu=0.35$, $\gamma=0.9$) \cite{att14}. These models also produce a DCF which is flat near zero time. However, the data collapse of the DCF with scaled time $t/\xi^z$ occurs for all times, not for the finite time interval observed in natural swarms \cite{cav17}. In addition, the qualitative shape of the observed swarms is quite different from the uniform density phase expected near an ordering transition of the ISMs. Thus, the ISM theories cannot describe the midge swarming observed in field studies.

\subsection{Effective one-particle theory}
Swarms in laboratory experiments consist of midges that are on average very weakly coupled, and yet tightly bound to the swarm itself \cite{ni15}. Consequently, Reynolds {\em et al} have proposed a model of an effective particle in a swarm velocity-dependent potential according to the Fokker-Planck equation \cite{rey17,vaa19}
\begin{subequations}\label{eq9}
\begin{eqnarray}
&&\frac{\partial P}{\partial t}+\mathbf{v}\cdot\nabla_xP +\nabla_v\cdot\!\left[\left(-\frac{\mathbf{v}}{T} + \langle \mathbf{A}|\mathbf{v},\mathbf{x}\rangle\right)\! P\right]\! = \frac{\sigma^2_v}{T}\nabla_v^2P, \label{eq9a}\\
&& \langle\mathbf{A}|\mathbf{v},\mathbf{x}\rangle= -\frac{\mathbf{v}\cdot(\mathbf{x}-\mathbf{x}_c)}{\sigma_r^2 |\mathbf{v}|^2} \mathbf{v}\times \left\{\begin{array}{cc} 3\sigma_v^2 &\mbox{(Gaussian velocities)}\\ 2\overline{s}(s+\overline{s}) &\mbox{(exponentially distributed)}
\end{array}\right.,\quad s=|\mathbf{v}|. \label{eq9b}
\end{eqnarray}
Here $\langle\mathbf{A}|\mathbf{v},\mathbf{x}\rangle$ and $\overline{s}$ are the conditional mean acceleration and the mean speed, respectively. The conditional mean acceleration expresses the effective restoring force that binds individuals to the swarm whose center is $\mathbf{x}_c$. It is determined assuming a spherical probability density compatible with experimental data \cite{rey17}. The free parameters $T$, $\sigma_r$ and $\sigma_v$ are set to unity. For Gaussian velocities, the mean acceleration (i.e., the restoring force) increase linearly with the distance to the swarm center, which is reminiscent of Okubo's model \cite{oku86} and of Gorbonos {\em et al}'s self-gravitating model \cite{gor16}. 
\end{subequations}

The Langevin equations corresponding to Eqs.~\eqref{eq9} are
\begin{subequations}\label{eq10}
\begin{eqnarray}
&&d\mathbf{x}=\mathbf{v}\, dt,  \label{eq10a}\\
&& d\mathbf{v}=\!\left(\langle\mathbf{A}|\mathbf{v},\mathbf{x}\rangle-\frac{\mathbf{v}}{T}\right) dt+ \sqrt{\frac{2\sigma_v^2}{T}}\, d\mathbf{W}(t), \label{eq10b}
\end{eqnarray}
where $\mathbf{W}(t)$ is the usual Wiener process. For undisturbed swarms $\mathbf{x}_c$ can be the origin of coordinates. Assuming a motion $\mathbf{x}_c(t)$ (for example, a periodic oscillation about the origin with $\mathbf{x}(0)=\mathbf{0}$), disturbances can be modeled \cite{vaa20}. If $N$ equations \eqref{eq10} are numerically simulated with the same function $\mathbf{x}_c(t)$, the result is a swarm of individual particles that only interact with the motion of its center. The results of such a simulation confirm that the correlation length increases with $N$ due to disturbances and produce a dynamic critical exponent $z=1$ \cite{vaa20}. The DCF does not seem to be flat at the origin and the data collapse seems to occur for all scaled times; see Fig.~4(b) of Ref.~\onlinecite{vaa20}. Furthermore, the effective one-particles theory yields no static critical exponents and $z=1$ disagrees with observations of natural swarms \cite{cav23}. Thus, the effective one-particle theory cannot describe the midge swarming observed in field studies.
\end{subequations}

\subsection{Harmonically confined Vicsek model}
The VM 
may move aligned particles away to the borders of a finite box, which is avoided by imposing periodic boundary conditions. However, swarms observed in the laboratory seem to adopt their shape by experiencing a central force that pulls them together, as explained in Section \ref{sec:2}. Adding a spring force to Eq.~\eqref{eq7}, we obtain the harmonically confined Vicsek model (HCVM):
\begin{eqnarray}
&&\mathbf{x}_j(t+\Delta t)=\mathbf{x}_j(t)+ \Delta t\,\mathbf{v}_j(t+\Delta t),\quad j=1,\ldots,N,\nonumber\\
&& \mathbf{v}_j(t+\Delta t)=v  \mathcal{R}_\eta\!\left[\Theta\!\left(\sum_{|\mathbf{x}_l-\mathbf{x}_j|<r_1 R_0}\mathbf{v}_l(t)-\mathcal{B} \mathbf{x}_j\right)\!\right]\!. \label{eq11}
\end{eqnarray}
Here $\mathcal{B}$ is a diagonal matrix of positive spring constants, and positions and velocities take values on $\mathbb{R}^3$. In our previous works, we considered an isotropic matrix with spring constant $\beta_0$ \cite{gon23,gon24}. Here we want to study the anisotropic case in which the constant along the vertical axis is smaller than those on the horizontal axes. 

To figure out whether we can approximate these discrete time equations by differential equations, we use data from the observations of natural midges reported in the supplementary material of Refs.~\onlinecite{cav17,att14}. We measure times in units of $\Delta t=0.24$ s, lengths in units of the time-averaged nearest-neighbor distance of the $20120910\_A1$ swarm  in Table I \cite{cav17}, which is $r_1=4.68$ cm, and velocities in units of $r_1/\Delta t$, whereas $v=0.195$ m/s. The nondimensional version of Eq.~(\ref{eq11}) has $\Delta t=1$, $r_1=1$, speed $v_0$ and confinement parameter $\beta$ given by
\begin{eqnarray}
v_0=v\, \frac{\Delta t}{r_1}, \quad \beta=\beta_0 \Delta t. \label{eq12}
\end{eqnarray} 
For the example we have selected, $v_0=1$, whereas other entries in the same table produce order-one values of $v_0$ with average 0.53 that exhibit the same behavior \cite{gon23}. Thus, our HCVM describing midge swarms is far from the continuum limit $v_0\ll 1$. Cavagna {\em et al} consider a much smaller speed for their periodic VM, $v_0=0.05$, closer to the continuum limit where derivatives replace finite differences \cite{cav17}. 

The nondimensional equations of the HCVM are Eqs.~\eqref{eq11} with $\Delta t=1$, $r_1=1$. Discrete time equations may have quite different attractors than the same equations with derivatives replacing differences (e.g., the discrete logistic map has a period doubling route to chaos, whereas the corresponding differential equation has one stable and one unstable time independent solution). This is certainly the case of Eqs.~\eqref{eq11}. As we shall review in the next section, this discrete time system has a number of periodic, quasiperiodic and chaotic attractors and exhibits a scale free chaos phase transition \cite{gon23} with a region of extended criticality between lines in the noise-spring constant that collapses to $\beta=0$ as $N$ goes to infinity \cite{gon24}. This transition produces static and dynamic critical exponents close to those measured in natural swarms. Its DCF has the same partial data collapse only at small values of rescaled time and flatness at zero time as the DCF derived from measurements. That the DCF collapses only at small values of rescaled time reflects the different time scales present in multifractal chaotic attractors, unlike the single time scale describing the DCF in ordering phase transitions \cite{gon23}. We will also examine how anisotropy improves the agreement of calculated and measured static exponents, produces a more realistic shape of the swarm and it may explain the difference between observations of swarms in the laboratory and in nature.

\section{Results for the Harmonically confined Vicsek model}\label{sec:4}
There are a few clues from midge swarms that we need to consider. 
\begin{itemize}
\item The time step in the VM can be interpreted as the time insects take to be aware of their neighbors and decide to align with the mean direction they provide. As mentioned before, this time step is not small. 
\item There is a central force responsible for swarm formation and maintenance that we can mimic as a spring force \cite{kel13,pat22}.
\item The swarm comprises a condensed core of insects and a vapor of insects going in and out from it \cite{sin17}.
\item Swarms form at dawn and dusk over a marker on the floor.
\item Midges acoustically interact when their distances are sufficiently small \cite{att14plos,att14} and react collectively to external acoustic signals \cite{ni15}.
\item The trajectories of single midges in a swarm follow L\'evy flights \cite{rey16}.
\item Whatever attractor is responsible for swarm criticality in field studies, it has to be scale free $\xi\propto N^{1/3}$.
\item Absent external environmental effects (breeze, temperature gradients, fluid flow, \ldots), swarms exhibit only short range velocity correlations in the laboratory, much smaller than their size \cite{niou15}. When external perturbations are added, the correlation length is proportional to the size of the swarm \cite{vaa20}.
\end{itemize}

\subsection{Isotropic confinement}
Here we review a few results obtained with isotropic confinement \cite{gon23,gon24}. From the numerical solutions of Eqs.~\eqref{eq11}, we find different attractors influenced by noise. Chaotic attractors are characterized by a positive largest Lyapunov exponent (LLE) that typically increases with noise strength. The precise algorithms used to distinguish attractors and find the critical lines separating different phases are described in Ref.~\cite{gon23}. Fig.~\ref{fig1}(a) from Ref.~\cite{gon24} (reproduced below) shows the phase diagram of the HCVM on the confinement vs noise plane for $N=500$, $v_0=R_0=1$. There are regions of deterministic and noisy chaos, noisy period-$\sigma$ (NP$\sigma$, not shown), noisy quasiperiodic (NPQ) attractors, and mostly noise. See Ref.~\cite{gon23} for the technical definition of noisy chaos using scale-dependent Lyapunov exponents, calculations of the LLE and reconstruction of chaotic attractors from time series obtained from the numerical simulations of the HCVM.  

\begin{widetext}
\begin{center}
\begin{figure}[ht]
\begin{center}
\includegraphics[clip,width=16cm]{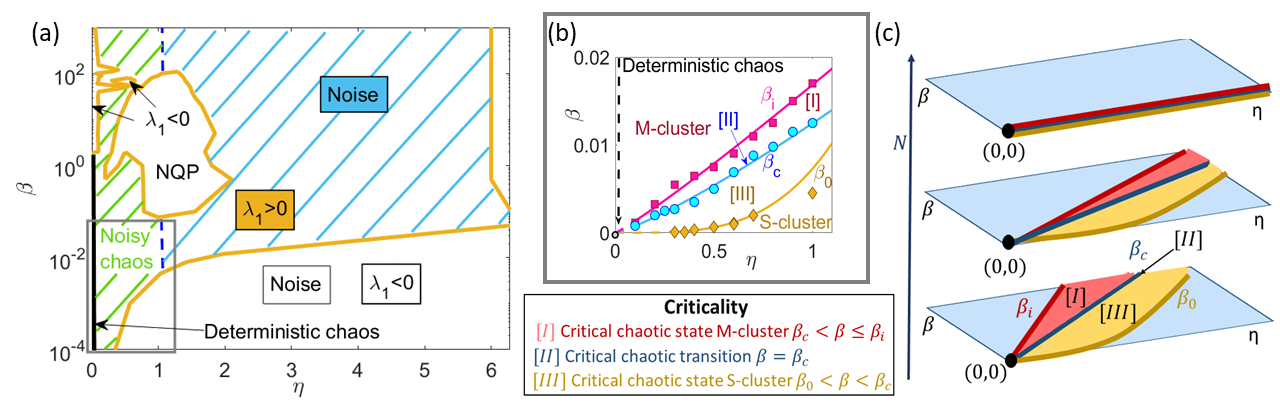}\\
\end{center}
\caption{Isotropic confinement. {\bf (a)} Phase diagram of the confinement vs noise plane indicating regions of deterministic and noisy chaos, noisy quasiperiodic (NPQ) attractors, and mostly noise. {\bf (b)} Regions [I]  $(\beta_c(\eta;N),\beta_i(\eta;N))$ (M-cluster or multicluster chaos), [III] $(\beta_0(\eta;N),\beta_c(\eta;N))$ (S-cluster or single-cluster chaos), and line [II] $\beta=\beta_c(\eta;N)$. In (a) and (b), $N=500$. {\bf (c)} Shrinking of the criticality region as $N$ increases. {\em Adapted from Fig.~1 of Ref.~\onlinecite{gon24}}. 
 \label{fig1}}
\end{figure}
\end{center}
\end{widetext}

The curve $\beta_0(N;\eta)$ separates chaotic and nonchaotic attractors for fixed $N$. On this curve, the LLE is zero and the correlation length is proportional to the size of the swarm. This indicates scale-free behavior and characterizes the scale-free-chaos phase transition in the limit as $N\to\infty$. Fig.~\ref{fig1}(b) shows different critical curves in the phase diagram. Besides $\beta_0(N;\eta)$, the correlation time reaches its minimum for fixed $N$ and $\eta$ at the curve $\beta_c(N;\eta)$. This curve separates the region of chaotic attractors where all particles belong to a single cluster from the region of multicluster chaotic attractors \cite{gon23}. The LLE is maximal on the curve $\beta_i(N;\eta)$. As $N\to\infty$ these three critical curves collapse to $\beta=0$ at the same rate as depicted in Fig.~\ref{fig1}(c) and the time averaged polarization also goes to zero. For finite $N$, there is a region of extended criticality between these curves \cite{gon24}. The correlation length and susceptibility are $\xi=$ argmax$Q(r)$ and $\chi=$ max$Q(r)$, respectively, where $Q(r)$ is the static cumulative correlation function, cf Eq.~\eqref{eq4}. These quantities can be related to $\beta_0$ and $\beta_c$ for different values of $N$ and the noise $\eta$, which produce the power laws $r_{0j}\propto\beta_j^{-\nu}$ and $\chi\propto\beta_j^{-\gamma}$. The same static critical exponents $\nu$ and $\gamma$ can be obtaining by studying the limit as $\eta\to 0$, in which the following relations hold \cite{gon24}:
\begin{subequations}\label{eq13}
\begin{eqnarray}
&& r_{0j}=D_j \beta_j^{-\nu} \eta^{-p_j}=\frac{D_j }{C_j^\nu}N^\frac{1}{3} \eta^{-p_j - \nu m_j}, \quad j=0,c,\quad \label{eq13a}\\
&&\chi = Q(r_0)=  Q_j\beta_j^{-\gamma}\eta^{q_j} =  \frac{Q_j}{C_j^\gamma}N^\frac{\gamma}{3\nu}\eta^{q_j-\gamma m_j},     \label{eq13b}
\end{eqnarray}
Eqs.~\eqref{eq13a} and \eqref{eq13b} allows to calculate the static critical exponents by fixing $N$ and drawing the curves $\beta_j(N;\eta)$ ($j=0,c$) as functions of $\eta$ (for small values of $\eta$). Furthermore, there is a linear relation between rescaled versions of  the perception range $x$ and $\beta_0$ or $\beta_c$:
\begin{eqnarray}
&&\eta^{m_{x0}}x=A_0+B_0\beta_0\eta^{-m_0},\quad\mbox{i.e., } \beta_0=B_0^{-1} \eta^{m_0+m_{0x}} (x-x_c),\label{eq13c}\\
&&\eta^{m_{xc}}x=A_c-B_c\beta_c\eta^{-m_c}, \quad\mbox{i.e., } \beta_c=B_c^{-1} \eta^{m_c+m_{cx}}(x_c-x).\label{eq13d}
\end{eqnarray}
\end{subequations}
Here the critical perception range at zero confinement is $x_c(\eta)=A_j\eta^{-m_{xj}}$, $j=0,c$. The values of all the parameters in Eqs.~\eqref{eq13} for isotropic confinement are: $C_c=1.5\pm 0.2$, $D_c= 1.33 \pm 0.04$, $m_c= 1.20\pm 0.04$, $p_c\approx 0$ ($p_c + \nu m_c = 0.55\pm 0.03$), $q_c\approx\gamma m_c$, $C_0=0.92\pm 0.22$, $D_0= 0.76 \pm 0.08$, $m_0\sim m_c+a_2N^{-n_2}$, $a_2=2.36\pm 0.07$, $p_0= 1.24-\nu m_c - \nu a_2 N^{-n_2}=0.72-\nu a_2 N^{-n_2}$ ($p_0+\nu m_0= 1.24\pm 0.11$), $n_2=0.24\pm0.01$, $m_{cx}=0.50\pm 0.03$, $A_c=2.00\pm 0.02$, $B_c=13.00\pm0.03$, $m_{0x} = 1.6\pm 0.2$, $A_0=2.0 \pm 0.2$, $B_0=219.8\pm 0.2$ \cite{gon24}. Note that $r_{0c}<r_{00}$ as $\eta\to 0$ and fixed $N$. Thus, the correlation length increases as $\beta$ decreases from $\beta_c(N;\eta)$ to $\beta_0(N;\eta)$ at fixed $N$ and small noise. 

Eqs.~\eqref{eq13c} and \eqref{eq13d} show that the critical confinements are proportional to $|x-x_c|$ for a fixed noise strength. Then the static power laws can be written in terms of  $|x-x_c|$ as in Eq.~\eqref{eq5}. The link between the (confinement) control parameter of the phase transition and the perception range allows us to relate theory and measurements of natural swarms. It was found for the first time in Ref.~\cite{gon24}.

For the 3D isotropic HCVM, the calculated critical exponents are $\nu=0.43\pm 0.03$, $\gamma = 0.92\pm 0.13$, and $z=1.24\pm 0.11$ (LS fit) or $z=1.37\pm 0.10$ (RMA fit) \cite{gon24}. The susceptibility as $\eta\to 0$ is calculated in linear response theory; see Eqs. (9)-(11) in Ref.~\onlinecite{gon24}. To mimic the data from experimentally observed swarms (recall that they correspond to different times, days or years, insect numbers and even midge species \cite{att14,cav17,cav23}), we built a mixture of points for different $N$ and $\eta$ on the curves $\beta_0$ and $\beta_c$. All these points belong to the extended criticality region where power laws hold. This data mixture produced different values of $z$ when depending on whether the fitting is linear squares or reduced major axis.

\subsection{Anisotropic confinement}
As shown in previous works, the swarm often has spherical shape for isotropic confinement (not surprisingly). It is known that midge swarms have weaker vertical confinement as compared to lateral confinement. This accounts for their elongated shapes. Here we will consider the HCVM with the following diagonal confinement matrix having nonzero entries:
\begin{eqnarray}
\mathcal{B}_{11}=\mathcal{B}_{22}=\beta, \quad\mathcal{B}_{33}=\frac{\beta}{2}.\label{eq14}
\end{eqnarray}

\begin{center}
\begin{figure}[ht]
\begin{center}
\includegraphics[clip,width=0.9\linewidth]{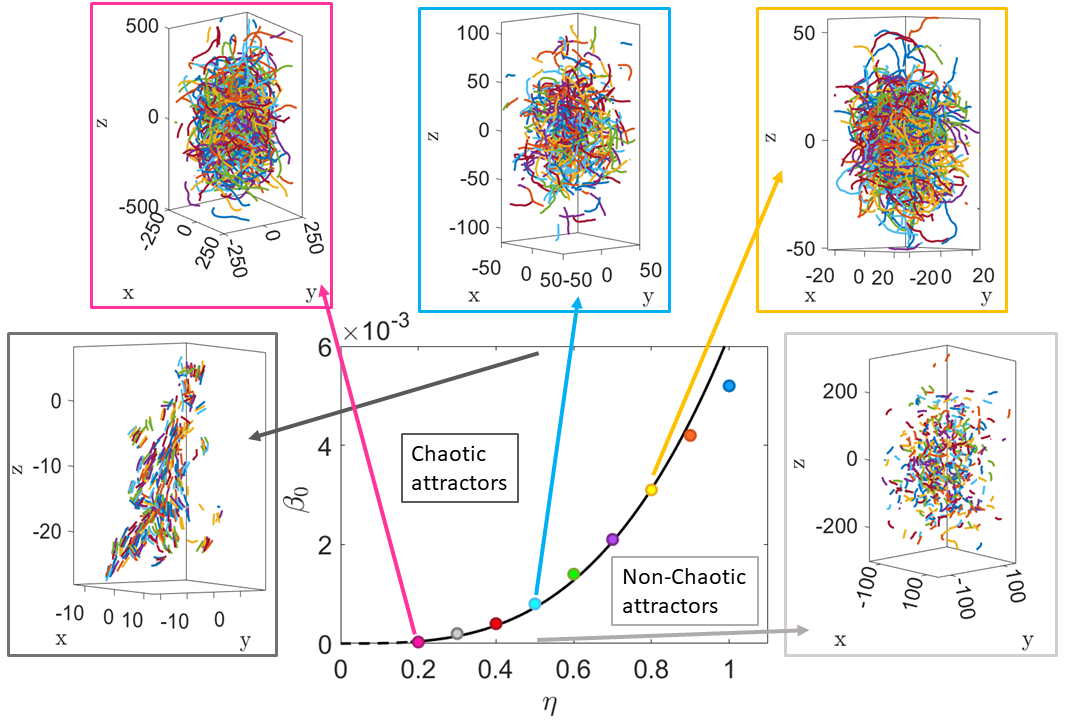}\\
\end{center}
\caption{Anisotropic confinement. Phase diagram in the $(\eta,\beta)$ plane. Chaotic and non-chaotic regions are separated by $\beta_0(\eta)$. The black curve is $\beta_0 = c_0 \eta^{m_0}$ with $m_0 = 3.10 \pm 0.13$ and $c_0 = 0.006 \pm 0.001$. Here $N=500$.}
 \label{fig2}
\end{figure}
\end{center}

We shall see that anisotropic confinement produces more realistic swarm shapes and improved critical exponents. Figure \ref{fig2} shows the critical curve $\beta_0(500;\eta)$ separating regions of chaotic and nonchaotic attractors in the phase diagram. The LLE is zero on this curve. Also shown are short time trajectories of the midges in the swarm for different confinement values along the critical curve, one swarm deep inside the chaos region and one swarm in the nonchaotic region. The swarms are elongated vertically and the particles outside the core are farther from it in the nonchaotic region, whereas the splitting into multicluster chaos becomes visible deep inside the chaotic region. The swarm occupies less volume as confinement increases. Video 1 in the Supplementary Material shows the motion of particles in the swarm for $\eta=0.5$ and critical confinement.

\begin{center}
\begin{figure}[ht]
\begin{center}
\includegraphics[width=0.3\linewidth]{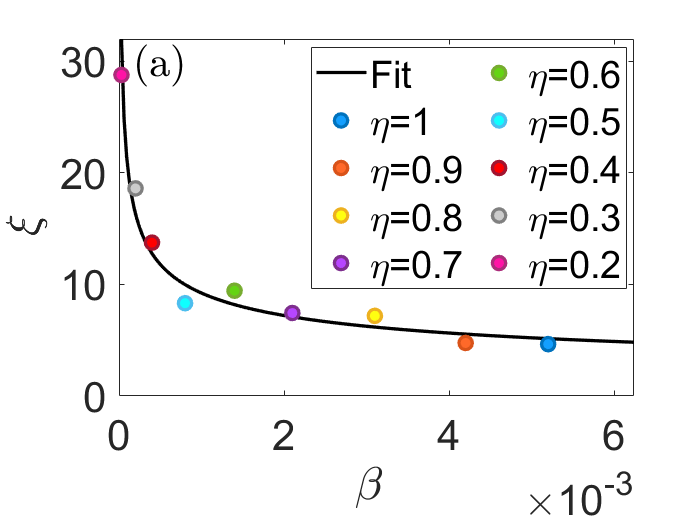}
\includegraphics[width=0.3\linewidth]{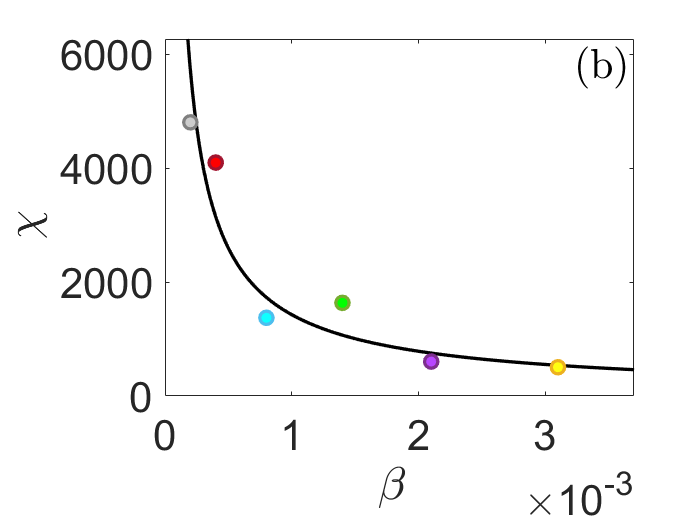}
\includegraphics[width=0.3\linewidth]{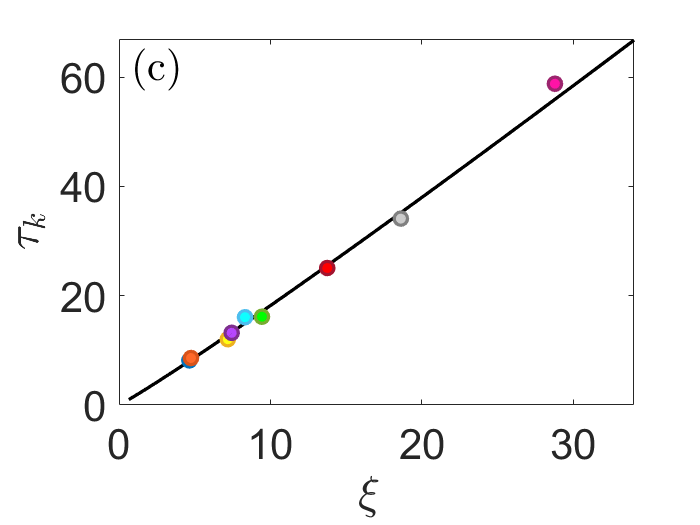}
\end{center}
\caption{ {\bf (a)} Correlation length and {\bf (b)} susceptibility as functions of $\beta$ for different values of the noise on the critical curve $\beta_0$. {\bf (c)} Determination of the dynamic critical exponent by LS fitting of correlation time versus correlation length. We find $\nu=0.35\pm 0.03$, $\gamma=0.86\pm 0.13$, $z=1.06\pm 0.03$ (LS fit). Here $N=500$.}
 \label{fig3}
\end{figure}
\end{center}

Drawing $\xi$, $\chi$ and $\tau_k$ in terms of $\beta_0$ for different noise values as shown in Fig.~\ref{fig3}, we can extract the static and dynamic critical exponents by using Eqs.~\eqref{eq13} as explained in Ref.~\cite{gon24}. Fixing $N=500$, we find $\nu=0.35\pm 0.03$, $\gamma=0.86\pm 0.13$, $z=1.06\pm 0.03$.

\begin{center}
\begin{figure}[ht]
\begin{center}
\includegraphics[clip,width=0.45\linewidth]{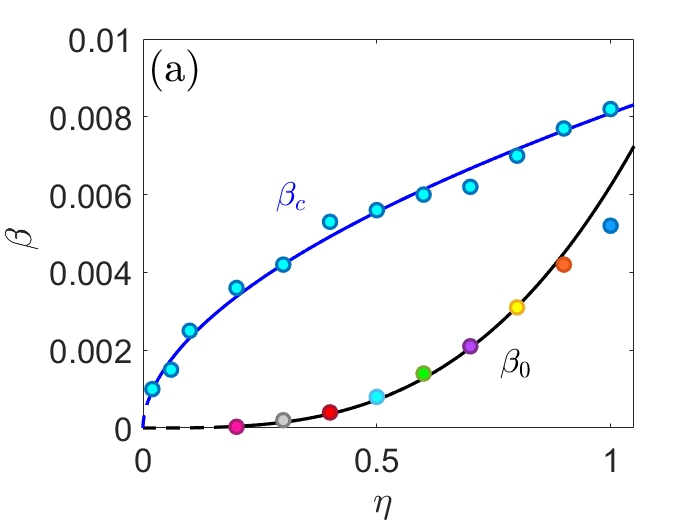}
\includegraphics[clip,width=0.45\linewidth]{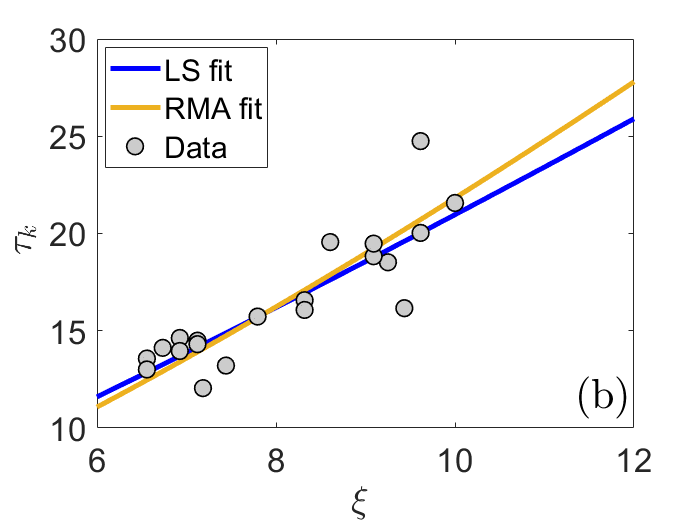}\\
\includegraphics[width=0.45\linewidth]{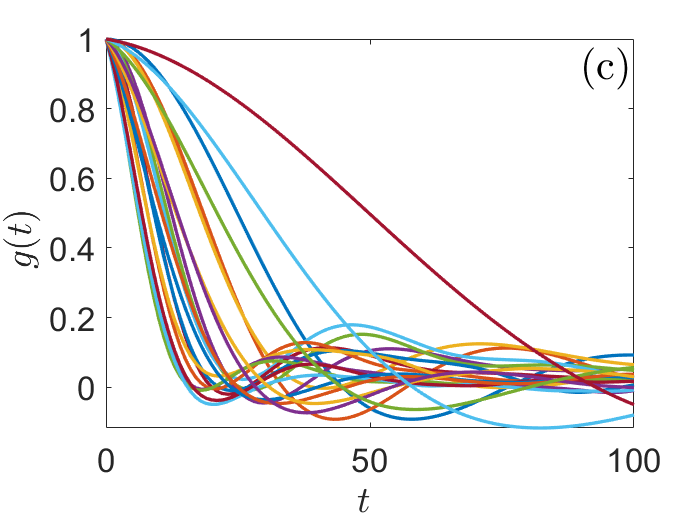}
\includegraphics[width=0.45\linewidth]{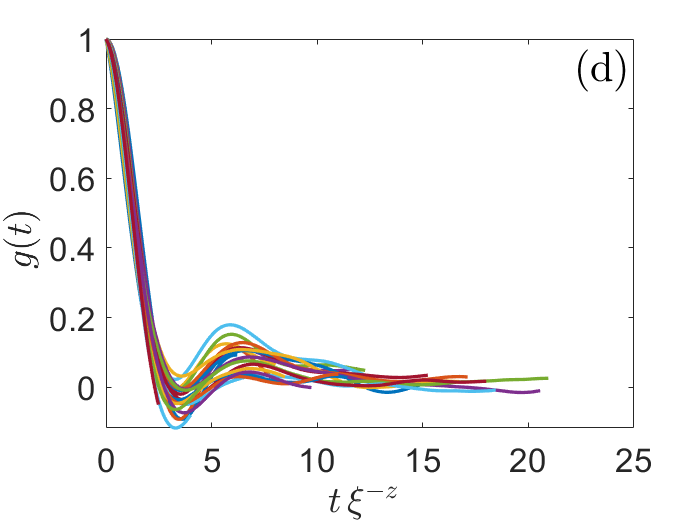}
\end{center}
\caption{ {\bf (a)} Critical region between $\beta_0$ and $\beta_c$ for $N=500$. The blue curve is $\beta_c = c_c\eta^{m_c}$, with $m_c= 0.54\pm 0.02$ and $c_c= 0.0081 \pm 0.0003$. {\bf (b)} LS and RMA fittings of the dynamic critical exponent for $\tau_k$ vs $\xi$ plane. We find $z_{LS}=1.17 \pm 0.15$, $z_{RMA}=1.33\pm0.15$. {\bf (c), (d)} Partial collapse of the DCF. The best visual collapse is with $z\approx1.1$. }
 \label{fig4}
\end{figure}
\end{center}

Next, we calculate the curve $\beta_c(500;\eta)$ of the anisotropic HCVM for small noise values. The result is shown in Fig.~\ref{fig4}(a). On this curve, we obtain the same critical exponents as before (LS fit) and a new power law for the LLE: $\lambda_1\propto\beta_c^\varphi$, with $\varphi=z\nu$ (the same relation holds for the isotropic HCVM \cite{gon23}). 

To mimic the data from natural swarms, we form a mixture of data on the region of extended criticality as follows. We choose points $(\eta,\beta)$ on the lines $\beta_0(500;\eta)$ and $\beta_c(500;\eta)$ in Fig.~\ref{fig4}(a), and numerically simulate eqs.~\eqref{eq11} with different random initial conditions. The critical exponent $z$ is determined from Eqs.~\eqref{eq6}(b) by fitting the resulting data using LS and RMA, as depicted in Fig.~\ref{fig4}(b). We obtain $z_{LS}=1.17 \pm 0.15$, $z_{RMA}=1.33\pm0.15$, which are close to the values obtained for the isotropic HCVM \cite{gon24}. Figures \ref{fig4}(c) and \ref{fig4}(d) illustrate the partial collapse of the DCF for small values of the scaled time $t\xi^{-z}$. 
As explained in Ref.~\onlinecite{gon23}, the correlation time as calculated in Eqs.~\eqref{eq6} captures only the longest time scale of the multifractal chaotic attractor. The latter has different time scales as a particle spends different times visiting different parts of the attractor, which accounts for the partial collapse of the DCF.

\begin{table}[ht]
\centering
\begin{tabular}{lccc}
\toprule
System & $\nu$ & $\gamma$ & $z$ \\
\hline
ISM \& RG 
& 0.748 & 1.171 & 1.35\\

effective one-particle
& - & - & 1\\

HCVM (mean field)
&0.5 & 1& 1\\

HCVM (2D, isotropic) 
& $0.30 \pm 0.02$ 
& $0.78 \pm 0.05$ 
& $0.99 \pm 0.03$ \\

HCVM (3D, isotropic) 
& $0.436 \pm 0.009$ 
& $0.92 \pm 0.05$ 
& $1.01 \pm 0.01$ \\

HCVM (3D, anisotropic) 
& $0.35 \pm 0.03$ 
& $0.86 \pm 0.13$ 
& $1.06 \pm 0.03$ \\

Experiments 
& $0.35 \pm 0.10$ 
& $0.9 \pm 0.2$ 
& $1.16 \pm 0.12$ \\
\hline
\end{tabular}
\caption{Comparison of static ($\nu$, $\gamma$) and dynamic ($z$) critical exponents for the isotropic HCVM in two and three dimensions, the anisotropic 3D HCVM (no data mixture), the mean-field HCVM, ISM, effective one-particle theory, and experimental measurements. All exponents are obtained from least-squares (LS) regression fits. RMA-fitted theoretical and experimental values of $z$ are given in the text.}
\label{table1}
\end{table}

The critical exponents obtained from mean-field \cite{gon23mf}, 2D \cite{gon2D}, isotropic 3D \cite{gon24}, and anisotropic HCVM (cf Fig.~\ref{fig3}) have been compiled in Table~\ref{table1}, which also includes those measured in natural swarms \cite{att14,cav23} and the other theories described in Section \ref{sec:2}. We observe that the static critical exponents obtained for the anisotropic 3D HCVM lie between the corresponding 2D and isotropic 3D values. Remarkably, both the static critical exponents $\nu$ and $\gamma$ of the anisotropic HCVM are in very good agreement with the experimental estimates within error bars. This suggests that anisotropic confinement effectively induces an intermediate critical behavior between 2D and 3D, providing a natural mechanism to reconcile theoretical predictions with experimental observations.

\section{Discussion: Solved case?}\label{sec:5}
Midge swarms have puzzling features: small polarization and, in the wild, strong correlations and scale free behavior with a correlation length $\xi$ proportional to the size of the swarm \cite{att14plos,att14}. In a lab environment without external perturbations, midges are tightly bound to the swarm, yet weakly coupled inside it, with small correlation length. Adding perturbations restores the scale free property \cite{vaa20}. Having scale free behavior, magnitudes related to correlation functions exhibit power laws with peculiar static and dynamic critical exponents \cite{att14,cav17,cav23}. Furthermore, the dynamic correlation function curves collapse only for short times when rescaled with a correlation time proportional to $\xi^z$. The same function is flat for short rescaled time. Finding a theory that explains these disparate features is akin to solving the whodunnit case of midge swarms.

In the previous sections, we have described different suspect theories and checked whether they are liable to solve the case thereby placing the observed power laws under firm ground \cite{stu12}. The main suspect is the harmonically confined Vicsek model which, when numerically solved, produces the critical exponents listed in Table \ref{table1} that compare well with those measured in natural swarms. Moreover, this model has the qualitative features of a flat DCF which displays partial collapse when written in terms of $t/\xi^z$. Let us discuss now loose ends.

Van der Vaart {\em et al} have shown that adding perturbations to midges in a lab setting restores the scale free behavior at least for the small swarm sizes they consider \cite{vaa20}. They back their observations by using the effective one-particle theory \cite{rey17,vaa19,vaa20}. In a way, the effective one-particle theory is reminiscent of the mean field HCVM \cite{gon23mf}. In both cases, a representative particle is subjected to the mean potential of a central force and different particles do not interact. Of course, the dynamics is different in these two theories. From the point of the isotropic HCVM, we would like to see whether there are lines in the phase diagram that separate strongly correlated states from weakly correlated ones as noise decreases. There are different scenarios to explore in future works. For the isotropic HCVM and small noise, we have seen that the correlation length $r_0$ increases as the confinement decreases to the critical line $\beta_0$ (recall $\beta_0=0$ for zero noise) separating chaotic and nonchaotic states. Assume a fixed confinement $\beta$ with a small correlation length for zero noise inside the chaotic region in Fig.~\ref{fig1} (corresponding to lab measurements). We could increase the noise until we arrive to the strongly correlated state such that $\beta=\beta_0(N;\eta)$ (corresponding to measurements of natural swarms). A second possibility is the phase transition from chaotic attractors to states of {\em flocking black holes} where infinitely many particles collapse to the same sites \cite{gon23}. We have only considered this phase transition as confinement and insect number go to infinity at fixed noise. However, it would be interesting to investigate whether there are critical lines related to this transition (nonzero to zero correlation length) that can be extrapolated to small noise and few particles. This could then be related to the question of whether lowering external perturbations produces a drop in the correlation length, as observed \cite{vaa20}. To this end, we know that the VM allows many particles to occupy simultaneously one position and, in fact, the mean field version of the HCVM consists of the same Eqs.~\eqref{eq11} for one particle ($N=1$) \cite{gon23mf}. One hurdle is that coalescing many particles in a single position may produce a large polarization (even unity if all the particles coalesce), contrary to the low ordering of midge swarms. At zero noise, we have found exact periodic and quasiperiodic solutions of the HCVM that can be polarized, partially polarized and unpolarized \cite{bon25}. There are also chaotic solutions at zero noise that may issue from those nonchaotic solutions \cite{bon25,gon23mf}. Whether these ideas are useful remains a task for the future.




Our findings related to the scale-free-chaos phase transition of the HCVM rest on numerical solutions and calculations that adapt the methodology used by Cavagna and coworkers to process the data from observations of natural swarms. We would like to have a theory of the scale-free-chaos phase transition that explains how the curve separating chaotic and nonchaotic states behaves as $N\to\infty$. For each $N$, the transition to chaos follows the quasiperiodic scenario \cite{ott93,cen10}, just as in the mean field case \cite{gon23mf}. There are renormalization group theories for this scenario but, unfortunately, restricted to very simple low dimensional maps such as the circle map \cite{lan86}. Extending them to the scale-free-chaos phase transition for infinitely many particles is a wide open problem. 
\bigskip




{\bf Acknowledgments.}\\ This work has been supported by the FEDER/Ministerio de Ciencia, Innovaci\'on y Universidades -- Agencia Estatal de Investigaci\'on (MCIN/ AEI/10.13039/501100011033) grant PID2024-155528OB-C22.


\begin{thebibliography}{}
\bibitem{bal08} M. Ballerini, N. Cabibbo, R. Candelier, A. Cavagna, E. Cisbani, I. Giardina, A. Orlandi, G. Parisi, A. Procaccini, M. Viale, V. Zdravkovic, Empirical investigation of starling flocks: a benchmark study in collective animal behaviour. Animal Behaviour {\bf 76}, 201-215 (2008).
\bibitem{gin15} F. Ginelli, F. Peruani, M.-H. Pillot, H. Chat\'e, G. Theraulaz, R. Bon, Intermittent collective dynamics emerge from conflicting imperatives in sheep herds. PNAS {\bf 112}(41), 12729-12734 (2015).
\bibitem{her11} J. E. Herbert-Read, A. Perna, R. P. Mann, T. M. Schaerf, D. J. T. Sumpter, A. J. W. Ward, Inferring the rules of interaction of shoaling fish. Proc. Natl Acad. Sci. USA {\bf 108}, 18726-18731 (2011).
\bibitem{kar14} I. Karamouzas, B. Skinner, S. J. Guy, Universal power law governing pedestrian interactions. Phys. Rev. Lett. {\bf 113}, 238701 (2014).
\bibitem{att14plos} A. Attanasi, A. Cavagna, L. Del Castello, I. Giardina, S. Melillo, L. Parisi, O. Pohl, B. Rossaro, E. Shen, E. Silvestri, M. Viale, Collective behaviour without collective order in wild swarms of midges. PLoS Comput. Biol. {\bf 10}(7), e1003697 (2014).
\bibitem{att14} A. Attanasi, A. Cavagna, L. Del Castello, I. Giardina, S. Melillo, L. Parisi, O. Pohl, B. Rossaro, E. Shen, E. Silvestri, M. Viale, Finite-size scaling as a way to probe near-criticality in natural swarms. Phys. Rev. Lett. {\bf 113}, 238102 (2014).
\bibitem{vaa20} K. van der Vaart, M. Sinhuber, A. M. Reynolds, N. T. Ouellette, Environmental perturbations induce correlations in midge swarms. J. R. Soc. Interface {\bf 17}, 20200018 (2020).
\bibitem{cav17} A. Cavagna, D. Conti, C. Creato, L. Del Castello, I. Giardina, T. S. Grigera, S. Melillo, L. Parisi, M. Viale, Dynamic scaling in natural swarms. Nat. Phys. {\bf 13}, 914-918 (2017).
\bibitem{mor11}T. Mora, W. Bialek, Are biological systems poised at criticality?  J. Stat. Phys. {\bf 144}, 268-302 (2011).
\bibitem{bia12} W. S. Bialek, Biophysics: Searching for Principles (Princeton University Press, Princeton, 2012).
\bibitem{stu12}M. P. H. Stumpf, M. A. Porter, Critical truths about power-laws. Science {\bf 335}, 665-666 (2012).
\bibitem{hua87}K. Huang, {\em Statistical Mechanics. 2nd ed} (Wiley, NY, 1987).
\bibitem{ami05}D. J. Amit, V. Martin-Mayor, {\em Field Theory, The Renormalization Group and Critical Phenomena, 3rd ed} (World Scientific, Singapore, 2005).
\bibitem{wil74}K. G. Wilson, J. Kogut, The renormalization group and the $\epsilon$ expansion. Phys. Rep. {\bf 12 C}, 75-199 (1974).
\bibitem{wil83} K. G. Wilson, The renormalization group and critical phenomena. Rev. Mod. Phys. {\bf 55}, 583-600 (1983).
\bibitem{hoh77}P. C. Hohenberg, B. I. Halperin, Theory of dynamic critical phenomena. Rev. Mod. Phys. {\bf 49}, 435-479 (1977).
\bibitem{dow55}J. A. Downes, Observations on the swarming flight and mating of Culicoides (Diptera: Ceratopogonidae). Trans. R. Entomol. Soc. London {\bf 106},  213-236 (1955).
\bibitem{oku74} A. Okubo, H. C. Chiang, An analysis of the kinematics of swarming of {\em Anarete Pritchardi} Kim (Diptera: Cecidomyiidae). Res. Popul. Ecol. {\bf 16}, 1-42 (1974).
\bibitem{oue22} N. T. Ouellette, A physics perspective on collective animal behavior. Phys. Biol. {\bf 19}, 021004 (2022).
\bibitem{pat22} M. L. Patel, N. T. Ouellette, Formation and dissolution of midge swarms. Phys. Rev. E {\bf 105}, 034601 (2022).
\bibitem{sin17} M. Sinhuber, N. T. Ouellette, Phase coexistence in insect swarms. Phys. Rev. Lett. {\bf 119}, 178003 (2017).
\bibitem{ni15} R. Ni, J. G. Puckett, E. R. Dufresne, N. T. Ouellette, Intrinsic fluctuations and driven response of insect swarms. Phys. Rev. Lett. {\bf 115}, 118104 (2015).
\bibitem{kel13} D. H. Kelley, N. T. Ouellette, Emergent dynamics of laboratory insect swarms. Sci. Rep. {\bf 3}, 1073 (2013).
\bibitem{rey16} A. M. Reynolds and N. T. Ouellette, Swarm dynamics may give rise to L\'evy flights. Sci. Rep. {\bf 6}, 30515 (2016).
\bibitem{niou15} R. Ni, N. T. Ouellette, Velocity correlations in laboratory insect swarms. Eur. Phys. J. Special Topics {\bf 224}, 3271-3277 (2015).
\bibitem{cav23} A. Cavagna, L. Di Carlo, I. Giardina, T. S. Grigera, S. Melillo, L. Parisi, G. Pisegna, M. Scandolo, Natural swarms in 3.99 dimensions. Nat. Phys. {\bf 19}, 1043-1049 (2023). 
\bibitem{gon23}R. Gonz\'alez-Albaladejo, A. Carpio, L. L. Bonilla, Scale free chaos in the confined Vicsek flocking model. Phys. Rev. E {\bf 107}, 014209 (2023). 
\bibitem{gon24}R. Gonz\'alez-Albaladejo, L. L. Bonilla, Power laws of natural swarms as fingerprints of an extended critical region. Phys. Rev. E {\bf 109}, 014611 (2024).
\bibitem{vic95}T. Vicsek, A. Czir\'ok, E. Ben-Jacob, I. Cohen, O. Shochet, Novel type of phase transition in a system of self-driven particles. Phys. Rev. Lett. {\bf 75}, 1226-1229 (1995).
\bibitem{vic12}T. Vicsek, A. Zafeiris, Collective motion. Phys. Rep. {\bf 517}, 71-140 (2012).
\bibitem{cha20} H. Chat\'e, Dry aligning dilute active matter. Ann. Rev. Cond. Matter Phys. {\bf 11}, 189-212 (2020).
\bibitem{cav15}A. Cavagna, L. Del Castello, I. Giardina, T. Grigera, A. Jelic, S. Melillo, T. Mora, L. Parisi, E. Silvestri, M. Viale, A. M. Walczak, Flocking and Turning: a New Model for Self-organized Collective Motion. J. Stat. Phys. {\bf 158}, 601-627 (2015).
\bibitem{cav24}A. Cavagna, J. Crist\'\i n, I. Giardina, T. S. Grigera, M. Veca, Discrete Laplacian thermostat for flocks and swarms: the fully conserved Inertial Spin Model. J. Phys. A {\bf 57}, 415002 (2024).
\bibitem{rey17}A. M. Reynolds, M. Sinhuber, N. T. Ouellette, Are midge swarms bound together by an effective velocity-dependent gravity? Eur. Phys. J. E {\bf 40}, 46 (2017). 
\bibitem{vaa19}K. van der Vaart, M. Sinhuber, A. M. Reynolds, N. T. Ouellette, Mechanical spectroscopy of insect swarms. Sci. Adv. {\bf 5}, eaaw9305 (2019).
\bibitem{oku86}A. Okubo, Dynamical aspects of animal grouping: Swarms, schools, flocks, and herds. Adv. Biophys. {\bf 22}, 1-94 (1986).
\bibitem{gor16} D. Gorbonos, R. Ianconescu, J. G. Puckett, R. Ni, N. T. Ouellette, N. S. Gov, Long-range acoustic interactions in insect swarms: an adaptive gravity model. New J. Phys. {\bf 18}, 073042 (2016).
\bibitem{gon23mf}R. Gonz\'alez-Albaladejo, L. L. Bonilla, Mean-field theory of chaotic insect swarms. Phys. Rev. E {\bf 107}, L062601 (2023). 
\bibitem{gon2D} R. Gonz\'alez-Albaladejo, L. L. Bonilla, Scale-free chaos in the 2D harmonically confined Vicsek model. Entropy {\bf 25}(12), 1644 (2023).
\bibitem{bon25}L. L. Bonilla, R. Gonz\'alez-Albaladejo, Exact solutions of the harmonically confined Vicsek model. Chaos, Solitons \& Fractals {\bf 191}, 115826 (2025).
\bibitem{ott93} E. Ott, {\em Chaos in dynamical systems} (Cambridge University Press, Cambridge UK 1993).
\bibitem{cen10} M. Cencini, F. Cecconi, A. Vulpiani, {\em Chaos. From simple models to complex systems} (World Scientific, New Jersey 2010). 
\bibitem{lan86}O. E. Lanford III, Renormalization group methods for circle mappings, 
in pp. 176-189 of {\em Statistical Mechanics and Field Theory: Mathematical Aspects},
ed. by T. C. Dorlas, N. M. Hugenholtz and M. Winnik (Springer-Verlag, Berlin 1986).
\end{thebibliography}
\end{document}